\documentclass[aps,prd,11pt]{revtex4}
\usepackage[colorlinks=true, pdfstartview=FitV, linkcolor=blue, citecolor=red, urlcolor=magenta]{hyperref}

\usepackage{graphicx}
\usepackage{latexsym}
\usepackage{amsmath}
\usepackage{amsfonts}
\usepackage{amssymb}
\usepackage{verbatim}

\newcommand{\be}{\begin{equation}}
\newcommand{\ee}{\end{equation}}
\newcommand{\ben}{\begin{eqnarray}}
\newcommand{\een}{\end{eqnarray}}

\begin{document}

\title{Induced higher-derivative massive gravity on a 2-brane  in 4D Minkowski space}
\author{D. Bazeia$^{a,b}$, F.A. Brito$^{a,b}$, and F.G. Costa$^{a,c}$ } 
\affiliation{
$^{a}$Departamento de F\'\i sica,
Universidade Federal da Para\'\i ba, Caixa Postal 5008,
58051-970 Jo\~ ao Pessoa, Para\'\i ba,
Brazil\\
$^{b}$Departamento de F\'\i sica,
Universidade Federal de Campina Grande, Caixa Postal 10071,
58109-970  Campina Grande, Para\'\i ba,
Brazil\\
$^{c}$Instituto Federal de Educa\c c\~ao, Ci\^encia e Tecnologia da Para\'\i ba (IFPB), Campus  Picu\'i, Brazil\\
}

\begin{abstract}
 In this paper we revisit the problem of localizing gravity in a 2-brane embedded in a 4D Minkowski space to address induction of high derivative massive gravity. We explore the structure of propagators to find well-behaved higher-derivative massive gravity induced on the brane. Exploring a special case in the generalized mass term of the graviton propagator we find a model of consistent higher order gravity with an additional unitary massive spin-2 particle and two massless particles: one spin-0 particle and one spin-1 particle. The condition for the absence of tachyons is satisfied for both `right' and `wrong' signs of the Einstein-Hilbert term on the 2-brane. We also find the Pauli-Fierz mass term added to the {\sl new massive gravity} in three dimensions and recover the low dimensional DGP model.
 
\end{abstract}\maketitle

\section{Introduction}
The interest on low dimensional gravity such as three-dimensional gravity has been renewed fairly recently \cite{nmg}.
The main interest of the present analysis is twofold. We consider two important aspects of gravity that are indeed connected by a special mechanism of gravity localization which take into account the fact of the graviton being a massless or massive particle. The first aspect has to do with low dimensional gravity that has been attracted much interest in the literature \cite{deser,mann1,callan,sikmann,mann}. This is justified because low dimensional gravity can bring us some new light about our difficulty of having a complete understanding of a quantized four-dimensional theory of gravity. The second concerns the inducing of gravity on a brane embedded in a higher-dimensional Minkowski space. This mechanism is completely different from localizing gravity via Kaluza-Klein with compact extra dimensions or through branes embedded in a warped higher-dimensional space as in Randall-Sundrum scenario \cite{rs}. The gravity localization with an embedded 3-brane in a 4D Minkowski space was first considered by Dvali-Gabadadze-Porrati (DGP) \cite{Dvali:2000hr} --- a similar set up with three 3-branes was also considered in \cite{Gregory:2000jc}. Since the Minkowski space has an infinite volume, one cannot find graviton zero mode. As the linearization of the equations of motion and in turn the structure of propagators show, the spectrum is a tower of continuum massive graviton states. However, there still appears four-dimensional gravity on certain scale. There is a natural crossover scale which allows that four-dimensional gravity can still be recovered if the distance between two masses on the 3-brane is much smaller than such a scale. On the other hand, the Newtonian potential modifies and assumes a 5D dimensional behavior for large distances compared with such a crossover scale. In a recent study we have considered an analogous set up, a 2-brane embedded into 4d Minkowski space to find lower-dimensional black hole solutions \cite{bbc1}. More recently a similar set up has been considered to address other issues \cite{peng} such as graviton localization and cascading gravity in lower dimensions. In this low dimensional DGP case four-dimensional gravity, as expected, is recovered at large distances on the 2-brane. On the other hand, as we shall see, the small distance behavior depends on the content of the induced gravity on the 2-brane. Indeed, as we shall see, the propagators for higher-derivative massive gravity seem to produce a new `cascading gravity' $4D\to3D\to2D$. The localization of gravity on a 2-brane is physically well-justified since we can see our real 3+1-dimensional universe as a stack of 2+1-dimensional branes \cite{Dai:2014roa}. Furthermore, this is also in the direction of the recently proposed ``vanishing dimension" scenario where at high energy (or short scales) the physics appears to be lower dimensional --- see \cite{Stojkovic:2014lha} for a review and references therein.

In this paper we revisit this problem of localizing gravity in a 2-brane embedded in a 4D Minkowski space to address induction of high derivative massive gravity. The paper is organized as follows. In Sec.~\ref{sec1} we briefly revise the DGP scenario for lower dimensions. In Sec.~\ref{sec2} we linearize the 4D action up to quadratic terms and integrated out on the fourth dimension to find the boundary (the induced 3D action) contribution. In Sec.~\ref{sec3} we study the structure of the propagators to identify the particle content of the theory. In Sec.~\ref{sec4}, since the `resonance massive graviton' can be generalized, we address the issue of graviton propagator with generalized mass term to find several scenarios with distinct particle content. Finally in Sec.~\ref{conclu} we make our final considerations.

\section{Induced gravity on a 2-brane}
\label{sec1}

In this section we consider the five-dimensional DGP scenario \cite{Dvali:2000hr} to extend our previous analogous set up in four-dimensions that we put forward in Ref.~\cite{bbc1}. The theory considers a 2-brane embedded in a flat four-dimensional spacetime (bulk). In this model the brane has invariants of the scalar curvature $R^{(3)}$ while  the bulk spacetime has a Ricci scalar term $R^{(4)}$. The full action can be separated into two pure (without matter fields) gravitational parts
\be S=S_{(4)}+S_{(3)},\ee
where
\be\label{act1} S_{(4)}=\frac{M^{2}_{4}}{2}\int d^{4}x\sqrt{|g|}R^{(4)}.\ee
and
\be\label{act2} S_{(3)}=\frac{M_{3}}{2}\int d^3x\sqrt{\left|q\right|}\left(\eta R^{(3)}+\gamma {(R^{(3)})}^2+\delta {R^{(3)}}^{\mu\nu}{R^{(3)}}_{\mu\nu}\right).\ee
This is a higher-derivative gravity model up to fourth order in derivatives \cite{stelle} in a three dimensional spacetime. The four dimensional line element is $ds^2_4=g_{ab}(x,w)dx^adx^b\ (a,b=0,1,2,4)$ and $ds^2_3=q_{\mu\nu}(x)dx^\mu dx^\nu\ (\mu,\nu=0,1,2)$ is the induced metric on the 2-brane at $x^4=w=0$ with $q_{\mu\nu}(x)\equiv g_{\mu\nu}(x,w\!=\!0)$. 

It follows that the induced scalar curvature $R^{(3)}$ is made out of this three-dimensional metric and has no dependence with the transversal fourth coordinate $w$. In order to study an effective action for gravity and its particle content we will adopt the same methodology of \cite{markus} for the traditional DGP case (see \cite{kurt} for a comprehensive recent review). 

It is convenient to write the action in terms of ADM-like variables \cite{adm}, $N=\sqrt{g^{44}}$ (lapse), $N_{\mu}=g_{4\mu}$ (shift) and the $4D$ metric $q_{\mu\nu}$ on the brane located at surfaces of constant $w=x^{4}$. The $4D$ part of the action can be written as
\be\label{actadm} S_{(4)}=\frac{M^{2}_{4}}{2}\int d^{4}xN\sqrt{\left|q\right|}[R^{(3)}+{(K^{(3)})}^2-{K^{(3)}}^{\mu\nu}{K_{(3)}}_{\mu\nu}].\ee
The $3D$ extrinsic curvature is given by
\be\label{extrinsic} K^{(3)}_{\mu\nu}=\frac{1}{2N}(q'_{\mu\nu}-\nabla_{\mu}N_{\nu}-\nabla_{\nu}N_{\mu}),\ee
where the prime means a derivative with respect to $w$.

The particle content of this theory is obtained expanding the action (\ref{actadm}) to linear order around the flat space. After integrating out the bulk we obtain an effective $3D$ action. Expanding the $4D$ graviton around flat space
\be g_{ab}=\eta_{ab}+H_{ab},\ee
and using the ADM-like variables with their expansions around flat space we have
\be\label{expan1} q_{\mu\nu}=\eta_{\mu\nu}+h_{\mu\nu},\;\;\;\;\;N_{\mu}=n_{\mu}, \;\;\;\;\;N=1+n.\ee
Furthermore, we have the following linear order relations
\be\label{expan2} H_{\mu\nu}=h_{\mu\nu},\;\;\;\;\;H_{\mu 4}=h_{\mu},\;\;\;\;\;H_{44}=2n.\ee
In order to obtain an effective $3D$ theory for the arbitrary brane boundary values we shall expand the action (\ref{actadm}) to quadratic order in $h_{\mu\nu}$, $n_{\mu}$, and $n$. We then solve the $4D$ equations of motion, subject to arbitrary boundary values on the brane and approaching zero as they tend to infinity. After doing that we plug this solution back into the action.

First, the bulk equations of motion are the vacuum Einstein equations at linear order
\be\label{bulk1} [R^{(4)}_{ab}]_{lin}=-\frac{1}{2}\square^{(4)}H_{ab}-\frac{1}{2}\partial_{a}\partial_{b} H+\frac{1}{2}\partial^{c}\partial_{a}H_{bc}+\frac{1}{2}\partial^{c}\partial_{b}H_{ac}=0.\ee
In the de Donder gauge,
\be\label{dedonder} \partial^{b}H_{ab}-\frac{1}{2}\partial_{a}H=0,\ee
Eq. (\ref{bulk1}) can simply be written as
\be\label{bulk2} \square^{(4)}H_{ab}=0.\ee

In terms of the ADM-like variables, Eq. (\ref{bulk2}) leads to three wave equations in the bulk for $h_{\mu\nu}$, $n_{\mu}$ and $n$,
\be \square h_{\mu\nu}+\partial^{2}_{w}h_{\mu\nu}=0,\ee
\be \square n_{\mu}+\partial^{2}_{w}n_{\mu}=0,\ee
\be \square n+\partial^{2}_{w}n=0,\ee
where $\square$ is the $3D$ Laplacian. These equations have the following solutions for $h_{\mu\nu}(x,w)$, $n_{\mu}(x,w)$ and $n(x,w)$, in terms of boundary values $h_{\mu\nu}(x)$, $n_{\mu}(x)$ and $n(x)$:
\be\label{sol1} h_{\mu\nu}(x,w)=e^{-w\Delta}h_{\mu\nu}(x),\ee
\be\label{sol2} n_{\mu}(x,w)=e^{-w\Delta}n_{\mu}(x),\ee
\be\label{sol3} n(x,w)=e^{-w\Delta}n(x).\ee
Here the operator $\Delta$ is the formal square root of the $3D$ Laplacian, $\Delta=\sqrt{\square}$.

Considering the $a=\mu$ and $a=5$ components of the gauge condition (\ref{dedonder}) we obtain
\be \partial^{\nu}h_{\mu\nu}-\frac{1}{2}\partial_{\mu}h+\partial_{w}n_{\mu}-\partial_{\mu}n=0,\ee
\be \partial^{\mu}n_{\mu}-\frac{1}{2}\partial_{w}h+\partial_{w}n=0.\ee
The boundary fields satisfy the following equations (at $w=0$):
\be\label{donder-brane1} \partial^{\mu}h_{\mu\nu}-\frac{1}{2}\partial_{\mu}h-\Delta n_{\mu}-\partial_{\mu}n=0,\ee
\be\label{donder-brane2} \partial^{\mu}n_{\mu}+\frac{1}{2}\Delta h-\Delta n=0.\ee

\section{Boundary effective action}
\label{sec2}

Let us now expand the $4D$ part of the action to quadratic order in $h_{\mu\nu}$, $n_\mu$, and $n$ and then plug in our solution.  Recall that this action given in (\ref{actadm}) reads
\be\label{act3} S_{(4)}=\frac{M^{2}_{4}}{2}\int d^{4}xN\sqrt{\left|q\right|}[R^{(3)}+{(K^{(3)})}^2-{K^{(3)}}^{\mu\nu}{K_{(3)}}_{\mu\nu}].\ee
We need to expand the $3D$ extrinsic curvature to first order, that is
\be\label{lin.extrin} K^{(3)}_{\mu\nu}=\frac{1}{2}(\partial_{w}h_{\mu\nu}-\partial_{\mu}n_{\nu}-\partial_{\nu}n_{\mu}).\ee
Thus, expanding (\ref{act3}), using (\ref{lin.extrin}) and integrating out by parts in $4D$ we have
\be\nonumber S_{(4)}=\frac{M^{2}_{4}}{2}\int d^{3}xdw\left[n\partial_{\mu}\partial_{\nu}h^{\mu\nu}-n\square h+\frac{1}{2}\partial_{\lambda}h_{\mu\nu}\partial^{\nu}h^{\mu\lambda}-\frac{1}{2}\partial_{\mu}h\partial_{\nu}h^{\mu\nu}-\partial_{w}h\partial_{\mu}n^{\mu}+\frac{1}{2}(\partial_{\mu}n^{\mu})^2\right.\ee
\be \left.+\partial_{w}h_{\mu\nu}\partial^{\mu}n^{\nu}+\frac{1}{2}n_{\mu}\square n^{\mu}\right]+\frac{M^{2}_{4}}{2}\int d^{3}x\left[-\frac{1}{4}h\partial_{w}h+\frac{1}{4}h_{\mu\nu}\partial_{w}h^{\mu\nu}\right].\ee
Now consider the following term into the action
\be S_{gf}=-\frac{M^{2}_{4}}{2}\int d^{4}x\left(\partial^{a}H_{ab}-\frac{1}{2}\partial_{a}H\right).\ee
This term does not contribute to the action because $4D$ equations of motion solve the de Donder gauge condition (\ref{dedonder}). Thus, we are allowed to add it. We can still write it in terms of the $3D$ variables
\be S_{gf}=\frac{M^{2}_{4}}{2}\int d^{3}xdw\left[-\frac{1}{2}\left(\partial^{\nu}h_{\mu\nu}-\frac{1}{2}\partial_{\mu}h+\partial_{w}n_{\mu}-\partial_{\mu}n\right)^2-\frac{1}{2}\left(\partial_{\mu}n^{\mu}-\frac{1}{2}\partial_{w}h+\partial_{w}n\right)^2\right].\ee
Plugging this into our $4D$ term the complete action can be now reduced to a boundary term at $w=0$, that is
\ben\nonumber S_{(4)}+S_{gf}=\frac{M^{2}_{4}}{2}\int d^3x\Big[-\frac{1}{2}h_{\mu\nu}\Delta h^{\mu\nu}+\frac{1}{4}h\Delta h-n\Delta n-n_{\mu}\Delta n^{\mu}+h\Delta h\nonumber\\
+n^{\mu}\left(-2\partial_{\mu}n-\partial_{\mu}h+2\partial^{\nu}h_{\mu\nu}\right)\Big].\een
Here $n^{\mu}$ plays the role of the vector St\"uckelberg field \cite{stuck} while $n$ plays the role of a gauge invariant auxiliary field. Fixing the gauge $n^{\mu}=0$ we can eliminate $n$ by using (\ref{donder-brane2}):
\be \Delta h=2\Delta n,\;\;\;\;\;h=2n.\ee

The resulting action has a Fierz-Pauli form \cite{fierz}, with an operator dependent mass term (resonance mass, or \textit{soft} mass) $m\Delta$ \cite{mass}:
\be\label{likefierz} S_{(4)}+S_{gf}=\frac{M_{3}}{2}\int d^3x\left[-\frac{1}{2}h_{\mu\nu}(m\Delta) h^{\mu\nu}+\frac{1}{2}h(m\Delta)h\right],\ee
where
\be m=\frac{M^{2}_{4}}{M_{3}},\ee
plays the role of the DGP or crossover scale. In the next section we shall linearize the $3D$ part to add it to the action (\ref{likefierz}) and get the full action for induced gravity on the 2-brane. We 
shall mainly focus on the propagator structure.

\section{The propagator structure of the full action on the 2-brane}
\label{sec3}

In order to obtain a full action on the 2-brane let us now address the issue of inducing higher-derivative gravity up to
fourth order derivatives in a three-dimensional spacetime described by the action (\ref{act2}), that is
\be S_{(3)}=\frac{M_{3}}{2}\int d^3x\sqrt{\left|q\right|}\left(\eta R^{(3)}+\gamma {(R^{(3)})}^2+\delta {R^{(3)}}^{\mu\nu}{R^{(3)}}_{\mu\nu}\right).\ee
As in the $4D$ part let us now expand the $3D$ action around a flat Minkowski background, i.e., $g_{\mu\nu}=\eta_{\mu\nu}+h_{\mu\nu}$, and keep only quadratic fluctuations. Then we express the action in terms of the Barnes-Rivers operators $P^{(2)}, P^{(1)}, P^{(0-s)}$ and so on.  
In the momentum space the set of three-dimensional operators in the corresponding Lagrangians is \cite{barnes}
\be\label{laghil} {\cal L_{\rm EH}}=\sqrt{g}\left(\eta\frac{M_{3}}{2}R^{(3)}\right)=\frac{1}{2}h_{\mu\nu}\left\{\eta\frac{ M_{3}}{4} k^2\left[P^{(2)}-P^{(0-s)}\right]^{\mu\nu,\alpha\beta}\right\}h_{\alpha\beta},\ee
\be\label{lagalfa} {\cal L_{\gamma}}=\sqrt{g}\left(\frac{\alpha}{2}{R^{(3)}}^2\right)=\frac{1}{2}h_{\mu\nu}\left\{2\gamma M_{3} k^2\left[P^{(0-s)}\right]^{\mu\nu,\alpha\beta}\right\}h_{\alpha\beta},\ee
\be\label{lagbeta} {\cal L_{\delta}}=\sqrt{g}\left(\frac{\beta}{2}{R^{(3)}}^{\mu\nu}{R^{(3)}}_{\mu\nu}\right)=\frac{1}{2}h_{\mu\nu}\left\{\frac{\delta M_{3}}{4}k^4\left[P^{(2)}+3P^{(0-s)}\right]^{\mu\nu,\alpha\beta}\right\}h_{\alpha\beta}.\ee

Now we must also express the Lagrangian related to (\ref{likefierz}) in terms of the spin projection operators. First we rewrite the Lagrangian in the form
\be {\cal L}_{(4)}+{\cal L}_{gf}=-\frac{1}{2}h_{\mu\nu}\left[M_{3}(m\Delta)(\eta^{\mu\alpha}\eta^{\nu\beta}-\eta^{\mu\nu}\eta^{\alpha\beta})\right]h_{\alpha\beta},\ee
such that
\be {\cal L}_{(4)}+{\cal L}_{gf}=\frac{1}{2}h_{\mu\nu}\left\{M_{3}(m\Delta)\left[-P^{(2)}-P^{(1)}+P^{(0-s)}+\sqrt{2}P^{(0-sw)}+\sqrt{2}P^{(0-ws)}\right]^{\mu\nu,\alpha\beta}\right\}\frac{1}{2}h_{\alpha\beta}.\ee
The spin projection operators in three spacetime dimensions $P^{(2)}$, $P^{(1)}$, $P^{(0-s)}$, $P^{(0-w)}$, $P^{(0-sw)}$, and $P^{(0-ws)}$ form a complete set and are defined as
\be P^{(2)}_{\mu\nu, \kappa\lambda}=\frac{1}{2}(\theta_{\mu\kappa}\theta_{\nu\lambda}+\theta_{\mu\lambda}\theta_{\nu\kappa}-\theta_{\mu\nu}\theta_{\kappa\lambda}),\ee
\be P^{(1)}_{\mu\nu, \kappa\lambda}=\frac{1}{2}(\theta_{\mu\kappa}\omega_{\nu\lambda}+\theta_{\mu\lambda}\omega_{\nu\kappa}+\theta_{\mu\nu}\omega_{\kappa\lambda}),\ee
\be P^{(0-s)}_{\mu\nu, \kappa\lambda}=\frac{1}{2}\theta_{\mu\nu}\theta_{\kappa\lambda},\;\; P^{(0-w)}_{\mu\nu, \kappa\lambda}=\frac{1}{2}\omega_{\mu\nu}\omega_{\kappa\lambda},\ee
\be P^{(0-sw)}_{\mu\nu, \kappa\lambda}=\frac{1}{\sqrt{2}}\theta_{\mu\nu}\omega_{\kappa\lambda},\;\;\;\;\; P^{(0-ws)}_{\mu\nu, \kappa\lambda}=\frac{1}{\sqrt{2}}\omega_{\mu\nu}\theta_{\kappa\lambda},\ee
In this context the transverse and the longitudinal operators ($\theta_{\mu\nu}$ and $\omega_{\mu\nu}$) are defined as
\be \theta_{\mu\kappa}=\eta_{\mu\nu}-\frac{k_{\mu}k_{\nu}}{k^2},\;\;\;\;\;\omega_{\mu\nu}=\frac{k_{\mu}k_{\nu}}{k^2},\ee
where the operators obey the following relationships
\be \theta_{\mu\rho}\theta^{\rho}_{\;\;\nu}=\theta_{\mu\nu},\;\;\;\;\;\omega_{\mu\rho}\omega^{\rho}_{\;\;\nu}=\omega_{\mu\nu}\;\;\;\;\;\theta_{\mu\rho}\omega^{\rho}_{\;\;\nu}=0.\ee

To write the Lagrangians (\ref{laghil}), (\ref{lagalfa}) and (\ref{lagbeta}) in terms of projection operators the following relationships were used
\be \left[P^{(2)}+P^{(1)}+P^{(0-s)}+P^{(0-w)}\right]_{\mu\nu, \kappa\lambda}=\frac{1}{2}(\eta_{\mu\kappa}\eta_{\nu\lambda}+\eta_{\mu\lambda}\eta_{\nu\kappa})\equiv I_{\mu\nu, \kappa\lambda},\ee
\be \left\{2P^{(0-s)}+P^{(0-w)}+\sqrt{2}[P^{(0-sw)}+P^{(0-ws)}]\right\}_{\mu\nu,\kappa\lambda}=\eta_{\mu\nu}\eta_{\kappa\lambda},\ee
\be [2P^{(1)}+4P^{(0-w)}]_{\mu\nu, \kappa\lambda}=\frac{1}{k^2}(\eta_{\mu\kappa}k_{\nu}k_{\lambda}+\eta_{\mu\lambda}k_{\nu}k_{\kappa}+\eta_{\nu\lambda}k_{\mu}k_{\kappa}+\eta_{\nu\kappa}k_{\mu}k_{\lambda}),\ee
\be {\sqrt{2}[P^{(0-sw)}+P^{(0-ws)}]+2P^{(0-w)}}_{\mu\nu, \kappa\lambda}=\frac{1}{k^2}(\eta_{\mu\nu}k_\kappa{}k_{\lambda}+\eta_{\kappa\lambda}k_{\mu}k_{\nu}),\ee
\be P^{(0-w)}_{\mu\nu, \kappa\lambda}=\frac{1}{k^2}(k_{\mu}k_{\nu}k_{\kappa}k_{\lambda}).\ee
The total gravitational Lagrangian ${\cal L}_{Grav}={\cal L}_{EH}+{\cal L}_{\gamma}+{\cal L}_{\delta}+{\cal L}_{(4)}+{\cal L}_{gf}$ can be expressed as
\be {\cal L}_{Grav}=\frac{1}{2}h_{\mu\nu}\textsl{O}^{\mu\nu,\alpha\beta} h_{\alpha\beta},\ee
where, in momentum space, we have
\be\nonumber \textsl{O}=\left[\delta\frac{M_{3}}{4}k^4+\eta\frac{M_{3}}{4} k^2-M_{3}mk\right]P^{(2)}-M_{3}mkP^{(1)}\ee
\be\nonumber +\left[\left(\frac{8\gamma+3\delta}{4}\right)M_{3}k^4-\eta\frac{M_{3}}{4}k^2+M_{3}mk\right]P^{(0-s)}\ee
\be+\sqrt{2}M_{3}mk(P^{(0-sw)}+P^{(0-ws)}).\ee
Now writing the $\textsl{O}$ operator in the form               
\be\textsl{O}=x_{2}P^{(2)}+x_{1}P^{(1)}+x_{s}P^{(0-s)}+x_{w}P^{(0-w)}+ x_{sw}P^{(0-sw)}+x_{ws}P^{(0-ws)},\ee
we can find the propagator 
\be \textsl{O}^{-1}=\frac{1}{x_{2}}P^{(2)}+\frac{1}{x_{1}}P^{(1)}+\frac{1}{x_{s}x_{w}-x_{sw}x_{ws}}(x_{w}P^{(0-s)}+x_{s}P^{(0-w)}-x_{sw}P^{(0-sw)}-x_{ws}P^{(0-ws)}),\ee
which can explicitly  be written in terms of the momentum as
\be\nonumber\textsl{O}^{-1}=\frac{1}{\delta\frac{M_{3}}{4}k^4+\eta\frac{M_{3}}{4} k^2-M_{3}mk}P^{(2)}-\frac{1}{M_{3}mk}P^{(1)}\ee
\be\nonumber +\frac{\left(8\gamma+3\delta\right)M_{3}k^4-\eta M_{3}k^2-4 M_{3}mk}{8M^{2}_{3}m^2k^2}P^{(0-w)}\ee
 \be\label{newpropag}+\frac{4\sqrt{2}M_{3}mk(P^{(0-sw)}+P^{(0-ws)})}{8M^{2}_{3}m^2k^2}.\ee
The tensor structure, as well as the behavior of the gravitational potential, for this propagator in the case $\gamma=\delta=0$ (simple Einstein-Hilbert term) was already investigated in \cite{peng}.  
In our case there is a fourth-order momenta in the propagator from where after integrating out on large momenta space (or equivalently for small distance $r$ compared with the crossover scale) this dominating term is expected to produce a linear gravitational potential $\sim r$ which 
is typical of a 2D spacetime. Thus in some sense the full propagator seems to produce a new `cascading gravity' $4D\to3D\to2D$.

\section{Graviton propagator with generalized mass term}
\label{sec4}

Replacing the mass term with an arbitrary function of the Laplacian, the resonance massive graviton can be generalized by \cite{gaba}, \cite{dvali}, \cite{dvali.hof}
\be\label{newmass} m^2\rightarrow m^2(\Box).\ee
Now the propagator (\ref{newpropag}) can be written as
\be\nonumber\textsl{O}^{-1}=\frac{1}{\delta\frac{M_{3}}{4}k^4+\eta\frac{M_{3}}{4} k^2-M_{3}m^2(\Box)}P^{(2)}-\frac{2}{2M_{3}m^2(\Box)}P^{(1)}\ee
\be\nonumber +\frac{\left(8\gamma+3\delta\right)M_{3}k^4-\eta M_{3}k^2-4 M_{3}m^2(\Box)}{8M^{2}_{3}(m^2(\Box))^2}P^{(0-w)}\ee
 \be\label{newpropag-1}+\frac{4\sqrt{2}M_{3}m^2(\Box)(P^{(0-sw)}+P^{(0-ws)})}{8M^{2}_{3}(m^2(\Box))^2}.\ee
It is not difficult to see from the structure of the propagators that there are indeed modifications of Newtonian dynamics on the brane at a large distances (small momenta). In this regime the mass term has a Taylor expansion given by
\be m^2(\Box)=L^{2(\alpha-1)}\Box^{\alpha},\ee
with $L$ being a length scale and $\alpha$ being a constant. In order to modify Newtonian dynamics at large scales the mass term should dominate over the two derivative kinetic terms, so we should have $\alpha<1$. On the other hand, there is the constraint that the spectral function should be positive definite, so that there are no ghosts. So we have a lower bound $\alpha\geq0$ \cite{dvali}. The standard DGP model corresponds to $\alpha=1/2$.
Making $\alpha=0$ ($m^2(\Box)=1/L^2$), $\eta=-1$, $\delta=1/M^2$ and $\gamma=-3/8M^2$ we have the particular case 
\be\nonumber\textsl{O}^{-1}=\frac{4}{M_{3}}\frac{1}{\left(\frac{1}{M^2}k^4-k^2-\frac{4}{L^2}\right)}P^{(2)}-\frac{L^2}{M_{3}}P^{(1)}\ee
\be\label{newpropag-2} -\frac{L^4}{8M_{3}}\left(k^2+\frac{4}{L^2}\right)P^{(0-w)}+\frac{\sqrt{2}L^2}{2M_{3}}(P^{(0-sw)}+P^{(0-ws)}).\ee
This is just the Pauli-Fierz mass term added to {\sl the new massive gravity} in three dimensions \cite{nmg}. We can see by (\ref{newpropag-2}) that there are
massive poles in the sector of spin-$2$ graviton modes. One is the unitary mode of positive mass and positive norm. The other is the ghost of tachyonic mass and negative norm. So both unitarity and causality are violated because of the existence of this tachyonic ghost \cite{oda}.

We are now interested in analyzing an unexplored case in literature. It concerns the regime in which $\alpha\rightarrow 1$. Thus, the generalized mass term is now given by
\be m^2(\Box)\approx \Box=k^2.\ee
The propagator in this case is written as
\be\textsl{O}^{-1}=\frac{4M^2}{M_{3}}\frac{1}{\left[k^2-(4-\eta)M^2\right]k^2}P^{(2)}-\frac{1}{M_{3}k^2}P^{(1)}-\frac{(4-\eta)}{M_{3}k^2}P^{(0-w)}+\frac{\sqrt{2}}{2M_{3}k^2}(P^{(0-sw)}+P^{(0-ws)}).\ee
Here $m_{2}^2=(4-\eta)M^2$ is the mass spin-$2$ mode. The condition for the absence of tachyons is satisfied for both `right' and `wrong' $\eta$ signs. As is well known, ghost does not show up in three dimensions even if there exists a propagator like $1/k^2(k^2-m^2)$ \cite{oda2}. 

Thus, we have a model of consistent higher order gravity with an additional unitary massive spin-$2$ particle and two massless particles: one spin-$0$ particle and one spin-$1$ particle:
\be\textsl{O}^{-1}=\frac{4}{(4-\eta)M_{3}}\frac{m^{2}_{2}}{\left(k^2-m^{2}_{2}\right)k^2}P^{(2)}-\frac{1}{M_{3}k^2}P^{(1)}-\frac{(4-\eta)}{M_{3}k^2}P^{(0-w)}+\frac{\sqrt{2}}{2M_{3}k^2}(P^{(0-sw)}+P^{(0-ws)}).\ee
The gravitational potential can be found by integrating out the 2-momenta $\bf k$ via Fourier transform.

\section{Conclusions}
\label{conclu}

In this paper we consider the problem of localizing gravity in a 2-brane embedded in a 4D Minkowski space to address induction of high derivative massive gravity. The structure of propagators shows a well-behaved higher-derivative massive gravity induced on the 2-brane. We consider a special case in the generalized mass term of the graviton propagator, which ends up with a consistent higher order gravity. The condition for the absence of tachyons is satisfied for both `right' and `wrong' signs of the Einstein-Hilbert term on the 2-brane. By properly choosing the parameters of the theory one can find the Pauli-Fierz mass term added to the new massive gravity \cite{nmg} and recover the low dimensional DGP model. It would be interesting to consider this study to pursue possible new aspects of  the following cascading gravity: $4D\to3D\to2D$.

\acknowledgements

We would like to thank to CNPq, PNPD-CAPES, PROCAD-NF/2009-CAPES for partial financial support.

\end{document}